\documentclass[useAMS,usenatbib]{mn2e}

\usepackage{epstopdf,graphicx,color}
\newcommand{\apj}{ApJ}				% "(the) Astrophysical Journal"
\newcommand{\apjl}{ApJL}			% "letter at ApJ"
\newcommand{\apjs}{ApJS}			% "ApJ Supplement Series"
\newcommand{\aap}{A\&A}				% "Astronomy and Astrophysics"
\newcommand{\mnras}{MNRAS}			% "Monthly Notices of the Royal Astronomical Society"
\title[Two Populations of transition discs?]{Two populations of transition discs?}
\author[James E. Owen \& Cathie J. Clarke]{James E. Owen$^{1}$\thanks{Email:jowen@cita.utoronto.ca} and Cathie J. Clarke$^{2}$\\
$^{1}$Canadian Institute for Theoretical Astrophysics, 60 St. George Street, Toronto, M5S 3H8, Canada.\\
$^{2}$Institute of Astronomy, Madingley Road, Cambridge, CB3 0DS, England.}
\begin{document}

\pagerange{\pageref{firstpage}--\pageref{lastpage}} \pubyear{2002}

\maketitle

\label{firstpage}

\begin{abstract}
We examine the distribution of transition discs as a function of mm
flux. We confirm that as expected in any model in which most primordial
discs turn into transition discs and in which mm flux declines with time,
transition discs have lower mm fluxes on average than primordial discs.
However, we find that the incidence of transition discs does not, as
expected, fall monotonically towards large mm fluxes and we investigate
the hypothesis that these mm bright transition discs may have
a distinct physical origin. We find that mm bright transition discs
occupy a separate region of parameter space. Transition discs in the bright mm sub-sample have systematically higher accretion rates than those in the faint mm sub-sample, along with being systematically weighted to earlier spectral types.

%We investigate the transition disc fraction as a function of mm flux and confirm the previously identified excess of transition discs at high mm flux. Furthermore, we identify a further excess of transition discs at low mm fluxes. The  pile up of  transition discs at function low and high mm flux can be interpreted as evidence for two distinct populations of transition discs. We have investigated the statistical properties of these two sub-samples under the assumption they are distinct. Transition discs in the sub-sample with a high mm flux have systematically higher accretion rates and inner hole radii than those in the low mm flux sub-sample, along with being systematically weighted to earlier spectral types. Furthermore, the statistics of these two sub-samples are distinct in the inner-hole size, accretion rate and spectral type distributions adding weight to the idea of two distinct populations. 
\end{abstract}
%\vspace{-0.5cm}
%\vspace{-0.5cm}
\section{Introduction}
Protoplanetary discs with apparent opacity deficits at Near-IR (NIR) wavelengths were first identified in early samples by \citet{strom89} and \cite{skrutskie90}. The lack of protoplanetary discs observed to be globally optically thin at all radii 
(e.g. \citealt{ercolano11}; but see also \citealt{currie11}), led to the interpretation that, if inner-hole objects indeed represent an evolutionary stage from a primordial, optically thick state to a disc-less
state, then disc dispersal must preferentially occur from the inside-out.  Under this assumption they were consequently christened `transition discs', and their rarity allowed the calculation of a disc clearing time-scale of around $10-20\%$   of a disc's primordial lifetime for both solar type stars (e.g. \citealt{kenyon95,luhman10}), and low mass stars (e.g. \citealt{ercolano09,ercolano11}). These transition disc observations have led to numerous models which attempt to explain their origin. One category of models explains this opacity deficit in terms of clearing of both gas and dust, for example by: planet formation (e.g. \citealt{armitage99}); photoevaporation (e.g. \citealt{clarke01}) and MRI driven winds (\citealt{suzuki09}). The ofther category of models instead posits only a deficit of dust opacity in the inner disc, resulting from: grain growth (e.g. \citealt{dullemond05}); dust filtration by planets (e.g. \citealt{rice06}) or photophoresis \citep{krauss07}. At this stage it is not clear observationally or theoretically whether one of these processes dominate, whether they are all occurring, or transition discs are caused by some yet unidentified mechanism. 

The observed sample of transition discs has now grown to a sufficiently large size that one can begin to compare the properties of these discs to the models. \citet{owen11,owen12} compared the predictions of X-ray photoevaporation to the sample of observed transition discs in the $R_{\rm hole}$ and $\dot{M}_*$ plane and identified that only approximately $\sim 50\%$ were consistent with being created through photoevaporation: namely, those with small holes $R_{\rm hole}<20 $AU and low accretion rates $\dot{M}_*<10^{-8}$ M$_\odot$ yr$^{-1}$. Furthermore, several other authors have used properties such as the surface density profile, presence of gas and/or dust in the inner hole, or disc mass to postulate the origin of individual transition discs (e.g. \citealt{cieza08,cieza10,alexander09,espaillat10}).%cieza10,alexander09,merin10,espaillat10}).  

Perhaps the most surprising recent development is that presented by \citet{andrews11}; looking at discs with large holes (and therefore imageable in the sub-mm) they calculated that $\sim 25\%$ of discs in the upper quartile of the sub-millimetre (mm) flux distribution for Class II discs may be identified as transition discs. This result is contrary to our inference that transition discs represent an evolutionary stage where discs are transitioning from a primordial to disc-less state at the end of their lifetimes, as mm flux is observed to decrease with age (e.g. \citealt{andrews05,andrews07}). In this scenario one would have naively expected a transition disc fraction that decreases monotonically with increasing mm flux. 

The analysis of statistically significant studies of observed transition disc samples has only recently become possible, allowing us to investigate their properties and identify possible sub-samples, taking care not to throw away the valuable information contained within upper-limits (using Survival statistics e.g. \citealt{feigelson85}). In this letter we use a large sample of transition discs (not just those which are imageable in the sub-mm) collected from the literature, to investigate the ratio of transition discs to primordial discs as a function of the full mm flux distribution. We can then compare this sample to both the \citep{andrews11} result and our naive theoretical assumptions as to what it should be (i.e. peaked a low mm flux and then monotonically decreasing). In Section~2 we describe the transition disc sample and look at its basic properties. In Section~3 we discuss the more detailed properties and possible sub-samples of the transition disc population and in Section~4 we draw our conclusions.   
\vspace{-0.8cm}
\section{Sample}
We have collected a large sample of transition discs from the literature that have measured mm fluxes; we choose to use 1.3mm fluxes instead of 880$\mu$m  fluxes as the former are more readily available for the collected data. In the case that only 880$\mu$m or 1.2mm measurements are available we use equations presented in \citet{cieza08} (used to calculate disc mass - based on \citealt{andrews05,andrews07}) to convert measurements into 1.3mm fluxes, and then we normalise all fluxes to a common distance of 140 pc using the {\it Spitzer} c2d distances. The transition discs used in the sample are taken from: \citet{andrews11,andrews12,brown09,calvet02,calvet05,cieza08,cieza10,espaillat07,espaillat10,hughes10,kim09,merin10,najita07}.

 Many of these articles discuss the same transition discs, so we take care to remove multiple instances of the same source. In addition we remove any `transition' disc which is caused by a known close stellar companion (e.g. CoKu-Tau 4 Ireland \& Kraus 2008). The total sample includes a total of 76 objects classified as transition discs by the various authors. For the 76 sources that make up our sample, 60 sources have available accretion rate and spectral type determinations; additionally 32 sources have had inner holes measured either through spectral type fitting and/or mm imaging. 

%\subsubsection{Notes on Individual Samples}\label{sec:cieza}
It is worth noting that \citet{cieza08,cieza10} do not fit their sources for inner holes, and they use a transition disc definition that is based on measurements of the {\it Spitzer} colours. However, \citet{cieza10} do list a `turn-off wavelength' representative of the dust temperature for the inner hole; thus we follow \citet{owen11,owen12} in turning these wavelengths into conservative upper limits. `Turn-off wavelengths' of 8$\mu$m and $<5.8\mu$m are assigned inner hole
radii of $<10$AU and $<5$AU, respectively. This increases the number of objects with any kind of inner hole measurement to 50 objects. We caution in the case of the four sources that overlap between \citet{cieza10} and \citet{merin10}, only two of these sources are identified as transition discs with inner holes by \citet{merin10}. Therefore, it is important to bear in mind that the sources classified by \citet{cieza08,cieza10} may not all be discs with inner holes, and the only definitive method of inner hole measurement is imaging. 
\vspace{-0.5cm}
\subsection{The millimetre distribution}
We use the transition disc sample to construct the Kaplan-Meier product-limit (KM) estimator (an estimate of the survival probability $P(x_i>X)$ e.g. \citealt{feigelson85}) for the mm distribution of transition discs. Furthermore, to compare the transition disc sample to a sample of primordial discs we use the sample of Taurus \citep{andrews05}; Ophiuchus\citep{andrews07}; Chameleon \citep{henning93} \& Lupus \citep{nuernberger97}  to construct the KM estimator for the mm distribution of class II (transition discs have been removed) objects.  Since our class II sample is exclusively from young (1-2Myr) clusters we further restrict our comparison to transition discs in the following young clusters: Taurus, Ophiuchus, Serpens, Chamelon \& Lupus, as the mm distribution of class II discs does appear to evolve with time \citep{lee11,mathews12}. This leaves us with 48 transition discs to compare with our class II sample, although we note the mm distribution of these 48 discs is statistically indistinguishable from the total sample (dashed line in Figure~1), thus we use the total sample later when looking at the properties of transition discs. 

\begin{figure}
\centering
\includegraphics[width=0.85\columnwidth]{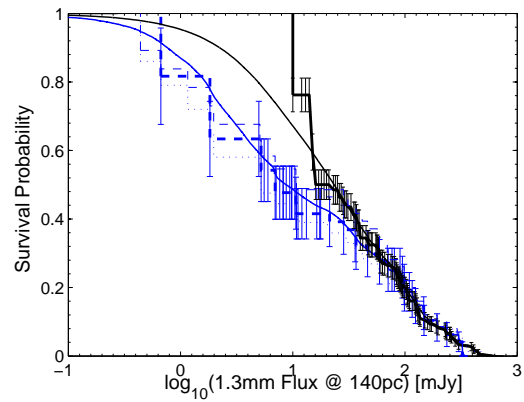}
\caption{The survival distribution for the mm fluxes of the primordial discs constructed
from the young class II objects, -- thick black line. 
Young transition discs are shown as the thick dashed blue line, while the dotted line shows the
same transition discs sample excluding the Cieza et al. objects, and the thin dashed line shows the total transition disc sample. The
error bars show the 1$\sigma$ errors in the KM estimator. The thin lines represent the smoothed distribution functions described in the text.}\label{fig:mmdist}
\end{figure}

We show the two survival distribution functions for the transition discs (thick dashed line) and primordial discs (thick solid line) with 1$\sigma$ errors in the KM estimator shown as error bars in Figure~\ref{fig:mmdist}.
\begin{figure}
\centering
\includegraphics[width=0.85\columnwidth]{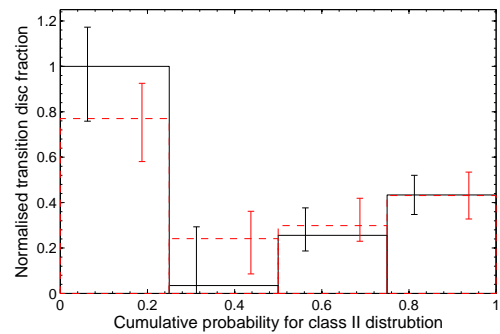}
\caption{Figure showing the relative ratio of transition discs to primordial discs in each quartile of the underlying primordial mm distribution, constructed from the KM estimator (solid) and smoothed distribution functions (dashed). The values are scaled such that the fraction in the lower quartile is unity for the KM estimator. Error-bars show representative errors estimated from the 1$\sigma$ errors in the KM/KS estimators.}\label{fig:tranfrac}
\end{figure} We note that removing the Cieza et al. transition discs {- all of which may not be discs with cleared inner holes, as their SED's were not fit for inner holes as discussed previously -} from the sample (dotted blue line in Figure~1) does not significantly affect the resulting survival distribution. { As survival statistics does not account for the fact astronomical measurements contain errors in measurements and upper limits \citep{feigelson85} which can mask an observational cut-off in a luminosity functions with true cut-off in a luminosity function, we compute smoothed distribution functions  by broadening each measurement or upper limit with an appropriate Gaussian or error function, allowing us to make sure a hard observational cut-off is not driving our conclusions seen in the data These smoothed distribution functions agree with those constructed from the KM estimator, indicating the use of exact upper limits in the survival analysis is not significantly affecting the result and our results are representative of the true luminosity functions. }

A glance at Figure~\ref{fig:mmdist} clearly shows that the mm distributions of primordial (class II) and transition discs are different, and the hypothesis that they come from the same underlying distribution can be quantitatively rejected at a level of 5.7$\sigma$ and 6.4$\sigma$ using the Gehan and logrank hypothesis tests, respectively. Furthermore, we can use the smoothed distribution functions to perform a Kolmogorov-Smirnov (KS) between the two populations, where we can reject the null hypothesis at the level of 2.4$\sigma$, { again indicating the use of survival statistics is appropriate for this data set.} Additionally, we can use these two distributions to construct the relative fraction of transition discs to primordial discs in each quartile of the class II mm distribution function, using both the KM estimator and the smoothed distribution function. This result is shown in Figure~\ref{fig:tranfrac}, where the relative fraction is scaled to unity in the lower quartile.

Figure~\ref{fig:tranfrac} shows a clear excess of transition discs in the lower quartile of the mm distribution, and a distribution function that is inconsistent with one that monotonically falls with increasing mm flux. Note that the transition disc and primordial samples do not come from the same star forming regions. If we sub-sampled the transition discs such that they are also in the come from the regions with complete class II mm samples, the sample size would be too low to be statistically useful. Therefore, Figure~\ref{fig:tranfrac} only contains information on the {\it relative} ratios of transition discs to primordial discs in different quartiles of the primordial disc mm flux distribution.

\vspace{-0.6cm}
\section{Discussion}
The high fraction of transition discs at large mm 
fluxes is difficult to explain with the canonical view of transition objects as discs caught in a final evolutionary stage before dispersal. In this picture one would expect a steeply decreasing transition disc fraction with increasing mm flux, as the latter also decreases with time. The high transition disc fraction at large mm flux contradicts this simple view and can be interpreted as evidence that either: (i) `transition discs' are a misnomer and these objects do not physically represent an evolutionary stage from disc-bearing and disc-less systems; or (ii) discs with inner holes can have multiple physical origins and the two populations shown in the data indicate two distinct classes of objects. The latter hypothesis will be further investigated in the following sections.
\vspace{-0.5cm}
\subsection{Spectral type distribution}
Perhaps the simplest interpretation of the observed distribution of transition
discs with mm-flux is the existence of two population of transition discs, one peaked at low mm 
fluxes which would be consistent with the classical picture of transition discs i.e. disc clearing, and a population of inner-hole objects with large mm 
fluxes which has a different physical origin. If we split the transition disc sample into two sub-samples about the
median Class II mm flux, 42 and 34 objects fall in the low- and high-mm flux sub-samples, respectively.   Discs with upper limits greater than the median value are placed in the sub-sample of discs with high mm fluxes, although we note this only represents a very small fraction (4 objects) of the total sub-sample. 

We can test the hypothesis that the mm faint
transition discs represent a later evolutionary phase in the life of
primordial discs by checking that they share the same distribution
of spectral types as the primordial disc population\footnote {This
test also assumes that there is no significant variation in
transition disc lifetime across the limited range of spectral types
($\sim$M to $\sim$G ) in the samples.}. The similarity in spectral type
distribution between the mm faint transition discs and the primordial
disc sample is demonstrated by the dashed and solid lines in
Figure 3: a KS test reveals that the two distributions have a high probability
of being drawn from the same population {(we can only reject the hypothesis they are different at $\sim0.6\sigma$)}. On the
other hand, the distribution of spectral types for the  mm bright
sub-sample of transition discs is clearly different from that of the
primordial population and the hypothesis that these
are drawn from the same population can be rejected at the $4.8 \sigma$
confidence level.  Clearly the bright mm sample contains a much higher
proportion of stars of earlier spectral type than the primordial (or  mm faint
transition disc) sample.

  This finding needs to be interpreted with care. The primordial
sample contains a large sample of objects from Taurus-Auriga ($\sim$ 40\%) which is well known (e.g. Goodwin et al. 2004) to be
deficient in stars of earlier spectral type. On the other hand, the
transition discs are a serendipitous sample culled from clouds at
a range of distances and there can be expected to be an obvious
over-representation of systems of earlier spectral type since these
are more luminous. However, it should be stressed that these
biases do not affect the obvious difference in spectral type distributions
between the mm bright \& faint transition discs since mm flux forms
no part of the object selection and identification.

  It is also worth noting that although the mm bright transition discs
contain a high fraction of earlier type stars, there is also a
significant population of mm bright discs at late spectral types
(see Figure~5). In fact, if we construct a figure
similar to Figure 2 but include only stars of spectral type later
than both $T_{\rm eff}=5000$ \& 4500K we obtain a qualitatively similar result, with a
concentration of sources in the lowest quartile and a flat or
gently rising distribution in the other quartiles. 

\begin{figure}
\centering
\includegraphics[width=0.85\columnwidth]{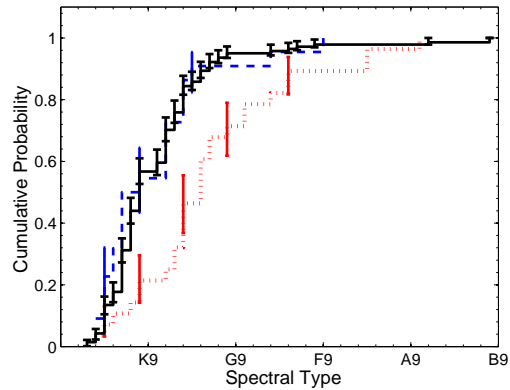}
\caption{Figure showing the distributions of spectral types for the primordial discs (solid), mm faint transition discs (dashed) and mm bright transition discs (dotted). The error bars indicate 1$\sigma$ errors.}\label{fig:spt}
\end{figure}   

\begin{figure}
\centering
\includegraphics[width=0.85\columnwidth]{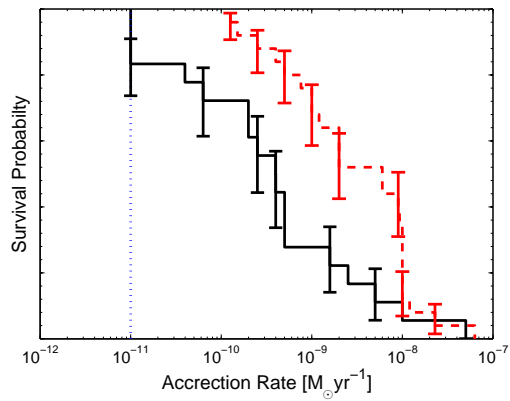}
\caption{The distribution function of accretion rates for mm bright transition discs (dashed) and faint mm fluxes (solid), the nominal accretion rate detection limit of $10^{-11}$ M$_\odot$ yr$^{-1}$ is shown as the dotted line.}\label{fig:hole_cdf}
\end{figure}

\vspace{-0.5cm}
\subsection{Transition disc properties}
We now examine the derived properties of transition disc in the two sub-samples. Figure~\ref{fig:hole_cdf} shows the distributions of accretion rates. It is clear that distributions of accretion rates are different for the two transition disc sub-samples, with the mm bright transition discs having considerable higher accretion rates. We confirm the difference in the distributions at the level of  $\sim 4.6\sigma$ and $\sim 5 \sigma$ for the accretion rates using the Gehan and logrank hypothesis tests, respectively. We do note that the mm bright sample does appear to have in general larger inner hole radii (see Figure~5) , although as the majority of inner holes are not measured or fit for the mm faint sample we do not draw any statistical conclusions. 

\begin{figure}
\centering
\includegraphics[width=0.85\columnwidth]{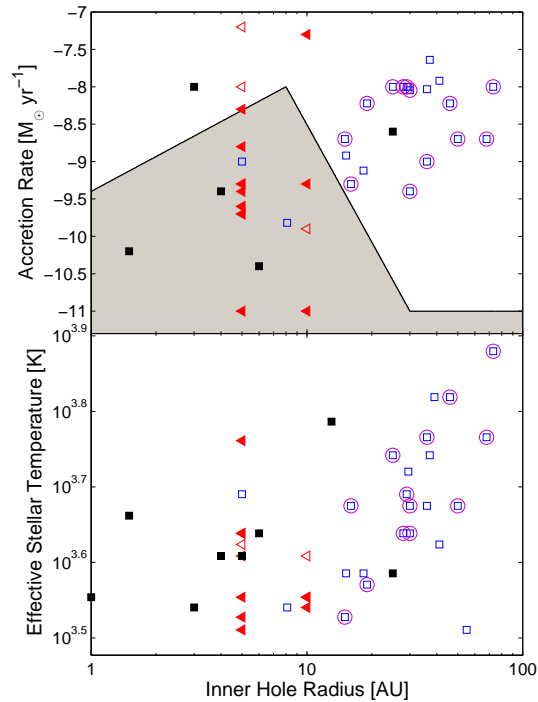}
\caption{Figure showing correlations between transition disc properties. The upper panel shows inner hole size and accretion rate, the lower panel shows inner holes size and effective stellar temperature. Triangles indicate the upper limits in inner hole radius, squares indicate actual measurements. Open symbols are for mm bright transition discs, and filled symbols are for mm faint transition discs. Objects at accretion rates of $10^{-11}$ M$_\odot$ yr$^{-1}$ are upper limits in accretion rates. We note some symbols contain multiple objects. The shaded region represents that predicted by photoevaporation; taken from \citet{owen12}- note the version of this Figure in \citet{owen12} contains more objects that do not have available mm fluxes. Points with directly imaged holes are also circled.}\label{fig:corr}
\end{figure}

Along with looking at the individual distribution functions for inner hole statistics we can look for correlations between the transition disc statistics and the parameter space they occupy in the individual sub-samples. It is of particular interest in order to test the models to study the distribution of observed transition discs in the $R_{\rm hole}$--$\dot{M}_*$ plane as shown in the
upper panel of Figure~\ref{fig:corr}. We see that, broadly speaking, the mm faint
objects (filled symbols) occupy the region of parameter space identified
by \citet{owen12} as being conceivably produced by photoevaporation whereas almost
all the bright mm objects (open symbols) lie outside this region. 
We also (lower panel) consider the relationship between inner hole
radius and stellar effective temperature in the two sub-samples.
We note that in both panels it is impossible to draw any conclusions
about correlations between these properties within the
faint mm sample on account of the  large number of
upper limits in hole size and accretion rate:
a large fraction of our faint mm sample is drawn from the sample of \citet{cieza10}
for which we have crudely estimated limits on the hole
size (see Section~2). 

%\begin{figure}
%\centering
%\includegraphics[width=0.9\columnwidth]{highmdotR.pdf}
%\caption{Same as Figure~\ref{fig:lowmdotR} but for the transition discs with high mm fluxes.}\label{fig:highmdotR}
%\end{figure}

However, correlations do appear in the mm bright transition discs, where there is clear evidence of a correlation between inner hole radius and accretion rate (we can reject the null hypothesis of no correlation at the level of 2.4$\sigma$), with large holes generally having higher accretion rates. Such a correlation was noted by \citet{kim09} with a much smaller sample of transition discs, and it is confirmed here. \citet{kim09} also noted correlations between accretion rate \& inner hole size with properties of the underlying star.  The lower panel
of Figure~5 shows clear evidence for a correlation between
inner hole radius and spectral type (a crude proxy for stellar mass)
in the mm bright sub-sample (we can reject the null hypothesis of no correlation at the level of 2.5$\sigma$). Moreover, it also clearly shows that the holes in
the
mm bright discs are systematically larger than those in the mm faint discs
{\it at all spectral types} and therefore the difference in inner hole
radii shown in the left hand panel of Figure~4 is not purely a result
of the mm bright discs being of earlier spectral type.  We note in passing that the correlations are still present (non-zero correlation coefficient) when one only considered those discs where the inner hole radius has been measured by imaging although it is less statistically significant where we can only reject the null hypothesis of no correlations at the 1.3$\sigma$ level in both cases. Finally we note that there is no evidence for correlations with mm flux for the inner hole radius, accretion rate and spectral type in the two individual sub-samples.

We speculate the interpretation of these results is that transition discs are not dominated by one physical mechanism, but there are perhaps two mechanisms at play. One producing transition discs in mm faint discs (likely associated with the end of the discs life) with low accretion rates and small holes, with the most likely origin being X-ray photoevaporation (Owen et al. 2011). The second mechanism may not be associated with the end of the disc's life at all, or the processes itself cause a high mm flux; the high accretion rates and presence of small dust grains inside the mm cavities (Dong et al. 2012) point to planet formation (e.g. Rice et al. 2006). However, we caution that much theoretical work must be done to
reconcile these observations with the planet formation scenario
since - in addition to reproducing the correlations suggested in this
work - there is considerable fine tuning required in the dust distributions
that can match both the SED and scattered light/mm images. 
\section{Summary \& Conclusions}
We have used a sample of transition discs collected from the
literature to look at their statistical properties, and compare their mm properties to those of primordial discs
(Class II objects) in Taurus Auriga. We demonstrate that transition discs are concentrated towards the lower end
of the mm flux distribution of primordial discs, as would be expected in
any model in which disc holes develop at a later evolutionary stage.
However, we find a significant minority of transition discs at all mm fluxes including very high mm fluxes similar to the incidence of transition discs in the
upper quartile of the primordial disc mm distribution noted by
\citet{andrews11}. 

We explore the hypothesis that the mm bright transition discs might
have a distinct physical origin by analysing the properties of 
two sub-samples of transition discs, split at the median of the
primordial disc mm distribution. Virtually all the mm bright transition discs have high
accretion rates and large inner holes.
The mm faint sub-sample has a range of accretion rates and systematically lower hole sizes
and broadly occupies the region of the accretion rate, hole size plane
that is consistent with X-ray photoevaporation models.  The two sub-samples
differ in accretion rate and inner hole size distributions at the $>
4 \sigma$ level. Furthermore, early type stars are over represented in the mm bright sample
compared to the primordial and faint mm transition discs. Nevertheless, the difference in properties cannot be
solely ascribed to differences in host stars. A flat tail of mm bright
sources are found at all spectral types and these have systematically
larger holes than mm faint objects of the same spectral type. Within the mm bright sub-sample there is
a strong correlation between inner hole size and spectral type (uncertainties
in hole sizes for many of the mm faint objects do not allow us to
assess this correlation within the mm faint sub-sample).

Finally, we ask whether there are any obvious selection effects that
drive the observed association between mm bright sources and large holes.
We believe there are no factors that would militate against the
identification of small holes in discs that are mm bright. Moreover,
since the evidence of even the largest inner holes in the sample is
based on spectral dips at $< 30 \mu$m, then the detection of such
a hole would not be undermined by a low flux at $1.3$mm. We therefore
do not believe that it is a detectability issue that is driving this
association. We thus need to ask a) why mm bright discs apparently
do not form small holes in small and large dust grains and b) is there a population of discs with such large holes
that they are classified as disc-less sources on the basis of their
near-infrared to mm SEDs? This latter issue was raised by Owen et al (2011)
although their subsequent work on Xray photoevaporation (Owen et al 2012)
suggested that the process of `thermal sweeping' should rapidly disperse
such low surface density relic discs. We also notes that some discs may manifest holes in mm images
while still showing high levels high levels of NIR excess as hinted by Andrews et al. (2011), whether one would class these discs within the same existing `transition' disc framework is unclear and certainly a matter for future debate.

We stress that these findings are necessarily suggestive rather than
conclusive and underline avenues of future investigation. Two obvious
shortcomings are i) the lack of fitted inner hole radii in a number
of transition discs in the literature and ii) the lack of good control
samples for the mm properties of primordial discs at a range of spectral
types. Clearly {\it ALMA} will assist in the latter respect.

\section*{Acknowledgments}
We thank the anonymous referee for comments which helped improve this work.
We thank Barbara Ercolano, Chris Thompson and Mike Irwin for useful discussions.  JO is grateful to hospitality from the IoA Cambridge, during the completion of this work.
\vspace{-0.5cm}
%\bibliographystyle{mn2e}
%\bibliography{JOCJrefs}

\end{document}